\documentclass[a4paper,12pt]{article}
\usepackage{mathrsfs}
\usepackage{graphicx}
\usepackage{amsfonts}

\begin{document}
\topmargin 0pt \oddsidemargin 0mm

\renewcommand{\thefootnote}{\fnsymbol{footnote}}
\begin{titlepage}
%\begin{flushright}
%hep-th/0206223
%\end{flushright}

\vspace{20mm}
\begin{center}
{\Large \bf A Dark Energy Model
 \\
 \vspace*{0.5cm}
 Characterized by the Age of the Universe}

 \vspace{20mm}
{\large
Rong-Gen Cai\footnote{Email address: cairg@itp.ac.cn}}\\
\vspace{5mm}
{ \em Institute of Theoretical Physics, Chinese Academy of Sciences, \\
   P.O. Box 2735, Beijing 100080, China}

\end{center}
\vspace{30mm} \centerline{{\bf{Abstract}}}
 \vspace{5mm}
 Quantum mechanics together with general relativity leads to the
  K\'arolyh\'azy relation and a corresponding energy density of
  quantum fluctuations of space-time.  Based on the energy density
  we propose a dark energy model, in which the age of the universe
  is introduced as the length measure. This dark energy is consistent with
  astronomical data if the unique numerical parameter in the dark
  energy model is taken to be a number of order one. The dark
  energy  behaves like a cosmological constant at early time and
  drives the universe to an eternally accelerated expansion with
  power-law form at late time. In addition, we point out a
  subtlety in this kind of dark energy model.

\end{titlepage}

\newpage
\renewcommand{\thefootnote}{\arabic{footnote}}
\setcounter{footnote}{0} \setcounter{page}{2}

%%=====================================================

Needless to say, the cosmological constant problem is one of the
biggest challenges in theoretical physics~\cite{Wen}. Naive
estimation leads the cosmological constant to be of the Planck scale
$(10^{19}Gev)^4$; if SUSY breaks at $Tev$ scale, the cosmological
constant should be in the order $(1 Tev)^4$. The discovery of the
current accelerated expansion of the universe causes the problem to
be more difficult to solve~\cite{Sne}, which implies that the
cosmological constant is in the scale $(10^{-3}ev)^4$. There exists
a big hierarchy difference between the theoretical estimation and
observation value.  Since the cosmological constant is related to
the vacuum expectation value of some quantum fields and it can be
measured only through gravitational experiments. Therefore the
cosmological constant problem is essentially a problem in quantum
gravity. Although a completely successful quantum theory of gravity
is still not yet available, quantum mechanics together with general
relativity may shed some lights on this issue.

 General relativity tells us that any classical physical laws
 concerning space-time can be verified without any limit in
 accuracy. To make a measurement of space-time, one has to introduce
 an experiment device. However, there exists a well-known Heisenberg uncertainty
 relation in quantum mechanics. The Heisenberg uncertain relation
 combining with general relativity leads to a fundamental scale of
 microstructure of space-time: Planck length $l_p \sim 10^{-33}cm$.
 Following the line of quantum fluctuations of space-time,
 K\'arolyh\'azy and his collaborators~\cite{Karo} made an
 interesting observation concerning the distance measurement for
 Minkowski space-time through a light-clock {\it Gedankenexperiment}
 (see also \cite{Maz1}): The distance $t$ in Minkowski space-time
 cannot be known to a better accuracy than
 \begin{equation}
 \label{eq1}
 \delta t = \beta t^{2/3}_p t^{1/3},
 \end{equation}
 where $\beta$ is a numerical factor of order one,  $t_p$ is the
 reduced Planck time, and throughout this paper, we use the units $c=\hbar
 =k_b=1$, so that one has $l_p=t_p=1/m_p$ with $l_p$ and $m_p$ being the
 reduced Planck length and mass, respectively.

The K\'arolyh\'azy relation (\ref{eq1}) together with the
time-energy uncertainty relation enables one to estimate a quantum
energy density of the metric fluctuations of Minkowski
space-time~\cite{Maz1,Maz2}. With the relation (\ref{eq1}), a length
scale $t$ can be known with a maximal precision $\delta t$
determining a minimal detectable cell $\delta^3$ over a spatial
region $t^3$. Thus one is able to look at the microstructure of
space-time over a region $t^3$ by viewing the region as the one
consisting of cells $\delta t^3 \sim t_p^2t$. Therefore such a cell
$\delta t^3$ is the minimal detectable unit of space-time over a
given length scale $t$ and if the age of the space-time is $t$, its
existence due to the time-energy uncertainty relation cannot be
justified with energy smaller than $\sim t^{-1}$ . Hence, as a
result of the relation (\ref{eq1}), one can conclude that if the age
of the Minkowski space-time is $t$ over a spatial region with linear
size $t$ (determining the maximal observable patch) there exists a
minimal cell $\delta t^3$, the energy of the cell cannot be smaller
than~\cite{Maz2}
 \begin{equation}
  \label{eq2}
  E_{\delta t^3} \sim t^{-1},
  \end{equation}
  due to the time-energy uncertainty relation. And then the energy
  density of the metric fluctuations of the Minkowski space-time
  is~\cite{Maz2}
  \begin{equation}
  \label{eq3}
 \rho_q \sim \frac{E_{\delta t^3}}{\delta t^3} \sim \frac{1}{t_p^2
 t^2}.
 \end{equation}
 The existence of this energy is necessary  to ensure the stability
 of the background space-time against the metric fluctuations since
 the relation (\ref{eq1}) determines the maximal accuracy allowed by
 the nature~\cite{Maz2}. More recently Maziashvili~\cite{Maz2} has
 investigated the cosmological implications of the K\'arolyh\'azy
 relation (\ref{eq1}) and the energy density (\ref{eq3}) in the different
 stages of cosmological evolutions including inflation epoch,
 radiation, matter  and dark energy dominated phases, respectively.
 It was found that to be consistent, $\beta^3 \approx 32\pi/3$, $72
 \pi/12$, and $8\pi/3$ during the radiation, matter and dark energy
 dominated phases, respectively. The  K\'arolyh\'azy
 relation (\ref{eq1}) and the energy density (\ref{eq3}) have also
 been discovered independently in \cite{Naok,Ng}. At this stage, we would like to
 mention that there exist some controversies on the validness of the
K\'arolyh\'azy relation (\ref{eq1}) in literature, see, for example,
\cite{in1,in2} and references therein.

 Here some remarks are in order on the K\'arolyh\'azy
 relation (\ref{eq1}) and the energy density (\ref{eq3}). First,
 let us mention  that the K\'arolyh\'azy
 relation (\ref{eq1}) obeys the holographic black hole entropy
 bound~\cite{Maz2}: the relation (\ref{eq1}) gives a relation between
 an
 UV cutoff $\delta l$ and the length scale $l$ of a system,
 $\delta l \sim l_p^{2/3} l^{1/3}$; the
 system has entropy
 \begin{equation}
 \label{eq4}
 S \le \left (\frac{l}{\delta l}\right)^3 \sim \left(\frac{l}{l_p}\right)^2 \sim S_{\rm
 BH},
 \end{equation}
 which is less than the black hole entropy with horizon radius
 $l$. Therefore, the K\'arolyh\'azy relation (\ref{eq1}) is a reflection
 of entanglement between UV scale and IR scale in effective quantum field
 theory~\cite{Cohen}.

 Second, the authors of \cite{Cohen}
 argued that considered the effect of gravity,
 the vacuum energy density $\rho_{\Lambda}$
 of a certain effective quantum
 field in a finite region with length scale $l$ cannot be
 arbitrary large, otherwise the region will collapse to a black
 hole with size $l$. This implies that  $\rho_{\Lambda} l^3 \le l/l_p^2$, which leads
 to
 \begin{equation}
 \label{eq5}
 \rho_{\Lambda} \sim \frac{1}{l_p^2 l^2}.
 \end{equation}
 One immediately sees that the energy density (\ref{eq3}) has the
 same form as the one (\ref{eq5}), the so-called holographic
 energy density, although the energy density (\ref{eq3})
 describes  quantum fluctuations of Minkowski space-time. The
 similarity between (\ref{eq3}) and (\ref{eq5}) might reveal some
 universal feature of quantum gravity since one arrives at
 (\ref{eq3}) and (\ref{eq5}) both by considering quantum effect of
 gravity, albeit in different way.

Third, let us mention that the cosmological implications of the
holographic energy (\ref{eq5}) has been investigated intensively.
Choosing the Hubble horizon $1/H$ of the universe as the length
scale $l$ in (\ref{eq5}), the holographic energy (\ref{eq5}) indeed
gives the observation value of dark energy in the universe. However,
as found by Hsu~\cite{Hsu}, in that case, the evolution of the dark
energy is the same as that of dark matter (dust matter), and
therefore it cannot drive the universe to accelerated expansion. The
same appears if one chooses the particle horizon of the universe as
the length scale $l$~\cite{Li}. An interesting proposal is made by
Li~\cite{Li}: Choosing the event horizon of the universe as the
length scale, the holographic dark energy (\ref{eq5}) not only gives
the observation value of dark energy in the universe, but also can
drive the universe to an accelerated expansion phase. In that case,
however, an obvious drawback concerning causality appears in this
proposal. Event horizon is a global concept of space-time; existence
of event horizon of the universe depends on future evolution of the
universe; and event horizon exists only for universe with forever
accelerated expansion. In addition, more recently, it has been
argued that this proposal might be in contradiction to the age of
some old high redshift objects, unless a lower Hubble parameter is
considered~\cite{Wei} (by the way, a complete list of references
concerning the holographic dark energy can be found in \cite{Wei}).

In this note we propose a dark energy model based on the energy
density (\ref{eq3}). The big difference from (\ref{eq5}) is that we
choose the age of the space-time as the length measure, instead of
the horizon distance of the universe. Thus the causality problem in
the holographic dark energy is avoided.  Note that energy density
for quantum fluctuations of matters in the universe has the same
order as the one (\ref{eq3}) for the metric fluctuation. We
introduce a numerical factor $n^2$ to parameterize some
uncertainties, for example, the species of quantum fields in the
universe, the effect of curved space-time (since the energy density
is derived for Minkowski space-time), etc. As a result, we write
down the energy density of quantum fluctuations in the universe as
\begin{equation}
\label{eq6} \rho_q = \frac{3n^2m_p^2}{T^2},
\end{equation}
as the dark energy in our universe, where $T$ is the age of the
universe, and the introduction of the number $3$ is for later
convenience.

The energy density (\ref{eq6}) with the current age of the
universe, $T\sim 1/H_0$ (here $H_0$ is the current Hubble
parameter of the universe), explicitly gives us the observed dark
energy density, provided the numerical factor $n$ is of order one
(Turn the logic around, the parameter $n$ can be estimated by
observational data in $\Lambda CDM$ model~\cite{WMAP}: $H_0=72\ km
\cdot s^{-1} \cdot Mpc^{-1}$, $T= 13.7 Gyr$ and $\Omega_{de}
=0.73$, then one has $n=1.15$) . Next, let us see whether the
energy density (\ref{eq6}) can drive the universe to accelerated
expansion. For the sake of simplicity, we first consider the case
without other matter in the universe. In this case, the Friedmann
equation for a flat FRW universe is
\begin{equation}
\label{eq7}
 H^2= \frac{1}{3m_p^2}\rho_q,
\end{equation}
where $H \equiv  \dot{a}/a$ is the Hubble parameter, $a$ is the
scale factor of the universe and the overdot stands for derivative
with respect to the cosmic time. The age of the universe can be
calculated through
\begin{equation}
\label{eq8}
 T=\int^a_0 \frac{da}{Ha}.
 \end{equation}
 Solving the Friedmann equation (\ref{eq7}) yields the evolution
 of the universe with the scale factor
 \begin{equation}
 \label{eq9}
 a= [n(H_0t +\alpha)]^n,
 \end{equation}
 where $\alpha$ is an integration constant, which can be
 determined by assuming the present scale factor $a_0=1$.
 We can see clearly from (\ref{eq9}) that the universe is in the
 accelerated expansion phase provided $n>1$. The equation of state
 for the energy density (\ref{eq6}) turns out to be
 \begin{equation}
 \label{eq10}
 w_q = -1 +\frac{2}{3n}.
 \end{equation}
 Indeed one can see from (\ref{eq10}) that the energy density can
 drive the universe to accelerated expansion if $n>1$. In addition, let us
 stress an interesting point here that without any inflaton, the energy
 density (\ref{eq6}) of quantum fluctuations can give rise to an
 inflationary period in the early universe.

 Now let us consider the case with dark (dust) matter in the
 universe.  In this case, the corresponding Friedmann equation is
 \begin{equation}
 \label{eq11}
 H^2 =\frac{1}{3m_p^2}(\rho_m +\rho_q).
 \end{equation}
 Defining the fraction energy density of dark matter as $\Omega_m=
 \rho_m/3m_p^2H^2$, and $\Omega_q =\rho_q/3m_p^2H^2$ for the dark
 energy, one has $\Omega_q= n^2/T^2H^2$. Using the Friedmann
 equation (\ref{eq11}), we get the equation of motion for
 $\Omega_q$ as
 \begin{equation}
 \label{eq12}
 \Omega_q'= \frac{2}{n}
 (\frac{3n}{2}-\sqrt{\Omega_q})(1-\Omega_q)\Omega_q,
 \end{equation}
 where the prime represents the derivative with respect to $\ln
 a$. This equation can be integrated analytically, one has
 \begin{eqnarray}
 \label{eq13}
 \frac{1}{n}\ln a + c_0  &=& -\frac{1}{3n-2}\ln (1-\sqrt{\Omega_q})
-\frac{1}{3n+2}\ln (1+\sqrt{\Omega_q}) \nonumber \\
   && + \frac{1}{3n} \ln \Omega_q +\frac{8}{3n(9n^2-4)}\ln
   (\frac{3n}{2}-\sqrt{\Omega_q}).
 \end{eqnarray}
 where $c_0$ is an integration constant, which can be determined
 by current observations, for example, WMAP with
 $\Omega_{q0}=0.73$ as $a=1$~\cite{WMAP}.  Although the expression
 (\ref{eq13}) is not instructive, it is easy to see that the fraction energy density
 $\Omega_q$ indeed decreases when it goes back to early time.
 To see the evolution behavior of
 the dark energy density, let us study its behavior in two different stages.
 The first one is the matter dominated phase, where $a \sim 0$ and $\Omega_q
 \sim 0$. In this case, we have the solution to the equation
 (\ref{eq12})
 \begin{equation}
 \label{eq14}
 \Omega_q \approx c_1 a^3,
 \end{equation}
 where $c_1$ is another integration constant. The fraction dark energy density increases during
 the epoch of matter domination. The other is the dark energy
 dominated phase, where $\Omega_m \sim 0$ and $\Omega_q \sim 1$.
 In that case, we get the solution to the equation (\ref{eq12})
 \begin{equation}
 \label{eq15}
 \Omega_q \approx 1-c_2 a^{-(3n-2)/n},
 \end{equation}
where $c_2$ is an integration constant. We see from (\ref{eq14}) and
(\ref{eq15}) that the fraction dark energy density increases quickly
and is independent of the parameter $n$ at earlier time of matter
dominated phase, while it approaches to one in a manner depending on
$n$ in the dark energy dominated phase at later time.

The equation of state for the dark energy can be easily obtained
through the formula, $w_q= -1 - \dot{\rho_q}/(3H\rho_q)$. It gives
us with
\begin{equation}
\label{eq16}
w_q= -1 +\frac{2}{3n}\sqrt{\Omega_q}.
\end{equation}
Once given the fraction energy density, the current equation of
state is completely determined by the parameter $n$. In Fig.~1 we
plot the current equation of state with respect to the parameter
$n$, provided $\Omega_{q0}=0.73$. We see that $w_{q0}\le -0.81$ as
$n\ge 3$. Therefore the equation of state is consistent with the
WMAP observation~\cite{WMAP}, as the parameter $n$ is taken to be
a number of order one.

\begin{figure}[htb]
\centering
\begin{minipage}[c]{.58\textwidth}
\centering
\includegraphics[width=\textwidth]{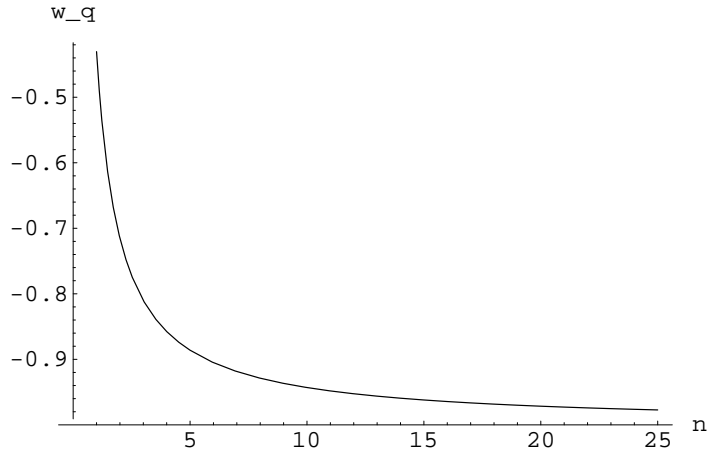}
\caption{This plot shows the current equation of state for the dark
energy versus the parameter $n$, provided $\Omega_{q0}=0.73$. }
\end{minipage}
\end{figure}

The equation of state (\ref{eq16}) has an interesting feature. At
 earlier time where $\Omega_q \to 0$, one has $w_q \to -1$.
Namely, the dark energy  behaves like a cosmological constant at
earlier time. At later time where $\Omega_q \to 1$, the equation of
state (\ref{eq16}) goes back to the case (\ref{eq10}). Therefore the
fate of our universe is an eternally accelerated expansion with
power-law form (\ref{eq9}) in this dark energy model.

After a close look at (\ref{eq14}) and (\ref{eq16}), a confusion
arises. In the matter dominated phase, if the dark energy is
negligible and $ a \sim t^{2/3}$, one then has $\Omega_q = 9n^2/4$.
This is obviously in contradiction to (\ref{eq14}). A more close
look at the equation (\ref{eq12}) tells us that this equation not
only holds for the form $T= \frac{n}{H\sqrt{\Omega_q}}$, but also
for another form $T'=T+\delta= \frac{n}{H\sqrt{\Omega_q}}$, where
$\delta$ is a constant, because the equation (\ref{eq12}) is
obtained by taking derivative of the form $T=
\frac{n}{H\sqrt{\Omega_q}}$ with respect to the cosmic time,
together with the energy conservation equation (or taking derivative
with respect to the cosmic time on both sides of equation
$\int^{a}_0 \frac{da}{Ha}=\frac{n}{H\sqrt{\Omega_q}}$). As a result,
the cosmic age $T'$ obtained by solving (\ref{eq12}) or using
(\ref{eq13}) might be different from the one (\ref{eq8}) by a
constant $\delta$. The constant $\delta$ can be determined by
$\delta=\frac{n}{H\sqrt{\Omega_q}}-\int^a_0 \frac{da}{Ha}$. Clearly
the constant $\delta$ depends on the parameter $n$ as well as the
current Hubble parameter $H_0$ and fraction energy density
$(\Omega_{m0}$) of dark matter (or equivalently, the fraction dark
energy density $\Omega_{q0}$)\footnote{A similar situation occurs
for the holographic dark energy~\cite{Li}. In that model, there
exist two expressions for the horizon distance: $R'_h=
\frac{c}{H\sqrt{\Omega_{de}}}$ and $ R_h=a \int^{\infty}_{a} \frac{d
a}{H a^2} $. The former can be obtained by solving a similar
equation as (\ref{eq12}) by adding the initial condition for the
current observational data. Nothing can guarantee these two
expressions are equal and a constant difference between them exists.
In fact, the same situation appears in the similar models.} . Thus
we can easily understand the results (\ref{eq12}) and (\ref{eq16}).
When $T \ll \delta$ at earlier time in the matter dominated phase,
the dark energy behaves as a cosmological constant, while it drives
the universe to an eternally accelerated expansion in a power-law
form (\ref{eq9}) at later time.

Another approach to way out the confusion is to consider $n$ in
(\ref{eq6}) as a slowly varying function of the age of the universe.
 For example, the
parameter $n$ might be dependent of the cosmic age in some way: in
the early stage, it changes the form (\ref{eq6}) in some manner so
that $n$ is negligible small (for instance, $n\sim T $) and at some
time, it approximately turns to be a constant. Indeed, the energy
form (\ref{eq3}) is derived from an argument in Minkowski spacetime.
In the early universe, where the space is highly curved and
dynamical, it is conceivable to think out that the parameter $n$
depends on the age of the universe in some way.

To summarize, we have proposed a dark energy model based on the
K\'arolyh\'azy relation (\ref{eq1}), and energy density of quantum
fluctuations of matter and metric in the universe. The dark energy
density (\ref{eq6}) has the same form as the holographic dark
energy, but we have introduced the age of the universe as the length
measure, instead of the horizon distance of the universe. Thus the
causality problem in the holographic dark energy is avoided. Our
dark energy model not only gives the observed value of dark energy
in the universe, but also can drive the universe to accelerated
expansion. Its equation of state can be consistent with astronomical
data, provided the unique parameter in the dark energy is taken to
be a number of order one. In this model, the dark energy
 behaves like a cosmological constant at early time and
it drives the universe to an eternally  accelerated expansion with
power-law form (\ref{eq9}) at later time.

\section*{Acknowledgments}
I would like to thank the referee for drawing my attention to
\cite{in1,in2}, H. Li, H. Wei, X. Wu, and Y. Zhang for useful
discussions. Also I am grateful to the comments appearing in M. Li's
blog (http://limiao.net/) on this dark energy model. The work was
supported in part by a grant from Chinese Academy of Sciences (No.
KJCX3-SYW-N2), and by NSFC under grants No.~10325525, No.~10525060
and No.~90403029.

\end{document}